# Optical Brewster metasurfaces exhibiting ultra-broadband reflectionless absorption and extreme angular-asymmetry

Huiying Fan[1], Jensen Li[2,*], Yun Lai[3,*], Jie Luo[1,*]

[1]School of Physical Science and Technology, Soochow University, Suzhou 215006, China

[2]Department of Physics, The Hong Kong University of Science and Technology, Clear Water Bay, Hong Kong, China

[3]National Laboratory of Solid State Microstructures, School of Physics, and Collaborative Innovation Center of Advanced Microstructures, Nanjing University, Nanjing 210093, China

*Corresponding authors: jensenli@ust.hk (Jensen Li); laiyun@nju.edu.cn (Yun Lai); luojie@suda.edu.cn (Jie Luo)

Impedance mismatch between free space and absorptive materials is a fundamental issue plaguing the pursue of high-efficiency light absorption. In this work, we design and numerically demonstrate a type of non-resonant impedance-matched optical metasurfaces exhibiting ultra-broadband reflectionless absorption based on anomalous Brewster effect, which are donated as optical Brewster metasurfaces here. Interestingly, such Brewster metasurfaces exhibit a unique type of extreme angular-asymmetry: a transition between perfect transparency and perfect absorption appears when the sign of the incident angle is changed. Such a remarkable phenomenon originates in the coexistence of traditional and anomalous Brewster effects. Guidelines of material selection based on an effective-medium description and strategies such as the integration of a metal back-reflector or folded metasurfaces are proposed to improve the absorption performance. Finally, a gradient optical Brewster metasurface exhibiting ultra-broadband and near-omnidirectional reflectionless absorption is demonstrated. Such high-efficiency asymmetric optical metasurfaces may find applications in optoelectrical and thermal devices like photodetectors, thermal emitters and photovoltaics.

## I. INTRODUCTION

Optical absorbers that exhibit ultra-broadband perfect absorption of light are promising for many applications such as photodetectors [1], thermal emitters [2] and photovoltaics [3], etc. In order to achieve perfect absorption, the incident light should be completely absorbed without any reflection or transmission. The transmission of light can be easily eliminated by increasing the thickness of the absorptive materials to a scale much larger than the wavelength. On the contrary, the elimination of the reflection caused by impedance mismatch at the air-absorber interface is challenging to achieve in an ultra-broad spectrum. This is because that the impedance of air is a fixed real value, while the impedance of absorptive materials is generally a frequency-dependent complex value, making the impedance matching in an ultra-broad spectrum an extremely difficult task.

The traditional method to mitigate the difficulty of an ultra-broadband impedance mismatch is based on surface topography [4, 5] like the absorbing wedges for microwaves, which can gradually vary the effective impedance from that of air to that of absorptive material. However, at optical frequencies, the scale of such wedges is reduced to micrometers and nanometers [6], which makes the fabrication more challenging. Another method is to introduce resonant structures to achieve destructive interference of the reflected light. The typical examples are the Dallenbach [7] and the Salisbury [8] screens. The recent advances of plasmonics, metamaterials and metasurfaces have provided efficient ways to manipulate light-matter interaction, making it possible to realize optically thin absorbers [9-18]. But due to the resonant nature, the absorption bandwidth is generally limited. In order to broaden the bandwidth, complex engineering of the dispersion [19, 20], multi-resonances at different wavelengths [21-29], and random plasmonic nano-structures [30-33] are usually required. In addition, angular-symmetric absorption is usually obtained, while the issue of absorption-related angular dependence or angular asymmetry have been rarely investigated until very recently [34-37].

Notably, ultra-broadband zero reflection and angular dependence both inherent exist in optics, which is known as the Brewster effect (BE) [38, 39]. For lossless dielectrics with almost constant permittivity, such as silica ($SiO_2$) and magnesium fluoride ($MgF_2$), zero reflection at the air-dielectric interface is easily achieved at the Brewster's angle over a broad spectrum for transverse-magnetic (TM) polarization. However, if loss is introduced to the dielectrics, the BE will be destroyed and reflection will appear. Due to this dilemma, although the BE has an inherently ultra-broad bandwidth, it has not been applied for ultra-broadband absorption until the anomalous Brewster effect (ABE) [40] was proposed very recently. The ABE establishes the ultra-broadband impedance matching between free space and anisotropic absorptive materials, thus bestowing reflectionless absorption of light in an unprecedented wide spectrum. Although microwave absorbers have been demonstrated to verify the concept and principle of the ABE [40], a design strategy for metasurfaces exhibiting ultra-broadband reflectionless absorption operating at visible and near-infrared wavelengths is currently lacking.

In this work, we analyze and numerically demonstrate a type of non-resonant impedance-matched metasurfaces exhibiting ultra-broadband reflectionless absorption based on the ABE, which are donated as optical Brewster metasurfaces here. Compared to traditional resonant optical absorbers [9-15], the Brewster metasurfaces exhibit a remarkable property, i.e. ultra-broadband extreme angular-asymmetry that varies from perfect transparency to perfect absorption. This rare effect is bestowed by the coexistence of the traditional BE and ABE, which occur correspondingly at the incident angles with opposite signs, due to protection by the principle of reciprocity. As shown schematically in Fig. 1(a), when TM-polarized light is incident from the left side under the Brewster's angle (i.e. $\theta_i = \theta_B = \arctan\sqrt{\varepsilon_d}$, where $\theta_i$, $\theta_B$ and $\varepsilon_d$ are, respectively, the incident angle, the Brewster's angle and the relative permittivity of the dielectric host), perfect transmission of light is achieved. On the other hand, when the incident light is flipped to the right side with $\theta_i = -\theta_B$, reflectionless light absorption is obtained instead. The angle $-\theta_B$ is denoted as the anomalous Brewster's angle.

The proposed optical Brewster metasurfaces are composed of periodic arrays of tilted metal films embedded in dielectric hosts. The material selection guidelines are systematically and quantitatively evaluated based on effective-medium description for visible and near-infrared wavelengths. It is found that the low-index dielectric materials like $SiO_2$ and $MgF_2$ are promising candidates for the dielectric hosts. On the other hand, the metals possessing a permittivity with a small real part and a large imaginary part, like titanium (Ti), vanadium (V), chromium (Cr), molybdenum (Mo) and Tungsten (W) are excellent candidates for the metal films. Numerical calculations show that near-perfect absorption can be obtained over the entire range of visible spectrum and even extends to the infrared regime. Moreover, we demonstrate that the ultra-broadband light absorption can be extended from the anomalous Brewster's angle $-\theta_B$ to the traditional Brewster's angle $\theta_B$ by using a metal back-reflector or a folded metasurface, so that the angular asymmetry is removed and the absorption performance is enhanced. Through engineering the tilt angle of metal films, a gradient optical Brewster metasurface exhibiting ultra-broadband and near-omnidirectional reflectionless absorption is demonstrated.

## II. OPTICAL BREWSTER METASURFACES AND THE UNDERLYING PHYSICS

The schematic graph of the proposed optical Brewster metasurface is shown in Fig. 1(a). It is composed of a periodic array of tilted metal films (relative permittivity $\varepsilon_m$, thickness $t$, tilt angle $\alpha$) aligned along the $y$ direction in a dielectric host (relative permittivity $\varepsilon_d$, thickness $d$). The separation distance between two adjacent metal films is $a$, which is much larger than $t$ (i.e., $a \gg t$), but smaller than free-space wavelength $\lambda_0$ to avoid diffractions.

We know that when TM-polarized light (magnetic fields along the $x$ direction) is incident onto the dielectric host in the absence of metal films, the reflection disappears under the Brewster's angle $\theta_B = \arctan\sqrt{\varepsilon_d}$ [38, 39]. Generally, the chromatic dispersion of conventional dielectrics like SiO$_2$ (~-0.035μm$^{-1}$) and MgF$_2$ (~-0.019μm$^{-1}$) is very weak in the optical regime [41], thus making it possible to obtain BE-induced zero reflection over the entire visible spectral range. In this case, the refracted light is normal to the direction of specular reflection, and the angle of refraction is $\theta_t = 90° - \theta_B$. Interestingly, when ultra-thin metal films are placed in parallel to the direction of the refracted light (i.e. $\alpha = \theta_t = 90° - \theta_B$), the BE won't be destroyed, as illustrated in the upper panel of Fig. 1(b). This is because the refracted light cannot "see" such ultra-thin metal films as its electric field $\mathbf{E}_t$ is perpendicular to these metal films [42], and therefore cannot be influenced by the metal films. As a result, the ultra-broadband zero reflection under the Brewster's angle $\theta_B$ preserves even in the presence of the metal films with $\alpha = 90° - \theta_B$.

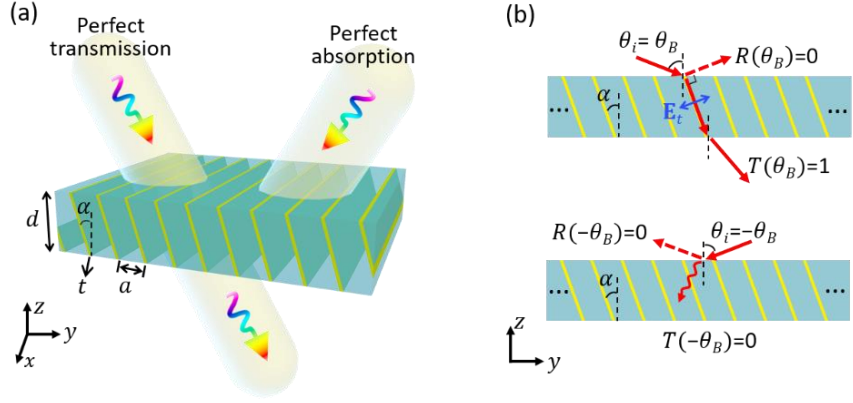

FIG. 1. (a) Schematic layout of an ultra-broadband reflectionless optical Brewster metasurface exhibiting extreme angular-asymmetry. The metasurface consists of a periodic array of metal films embedded in a dielectric host. (b) Upper panel: ultra-broadband perfect transparency due to the BE under the incident angle of $\theta_i = \theta_B$. Lower panel: ultra-broadband perfect absorption due to the ABE under $\theta_i = -\theta_B$.

It becomes interesting when the reciprocity principle [43] is applied. Now, we consider flipping the incident angle from $\theta_B$ to $-\theta_B$ (the lower panel of Fig. 1(b)). According to the reciprocity principle, the ultra-broadband zero reflection remains unchanged [43]. However, the electric field of the refracted light is no longer perpendicular to the metal films. In this case, currents would be induced along the metal films, leading to the dissipation of the refracted light. As a result, ultra-broadband reflectionless absorption of light under $\theta_t = -\theta_B$, as strictly protected by the reciprocity principle, can be obtained. The angle $-\theta_B$ thus can be denoted as the anomalous Brewster's angle, and the reflectionless phenomenon can be denoted as the ABE [40]. The optical metasurfaces exhibiting the ABE are donated as the optical Brewster metasurfaces.

Therefore, when switching the incident angle from the traditional Brewster's angle $\theta_B$ to the anomalous Brewster's angle $-\theta_B$, TM-polarized light will experience extreme angular-asymmetry from perfect transparency to perfect absorption in an ultra-wide spectrum. As a practical example, we take $SiO_2$ as the dielectric host, whose refractive index is approximately considered as a constant of 1.46 in the studied wavelength range 400-1400nm, thus the Brewster's angle is fixed at $\theta_B = 55.6°$. The metal films are made of Cr with $t = 10$nm, $a = 100$nm and $d = 400$nm. These Cr films are ultra-thin, satisfying the condition $t \ll a < \lambda_0$. The tilt angle is set as $\alpha = 90° - \theta_B = 34.4°$, such that there will be no reflection for light under $\theta_i = \pm\theta_B = \pm55.6°$ due to the BE and ABE.

Direct numerical proofs are presented in Fig. 2 using the software COMSOL Multiphysics. Here we consider TM-polarized light, whose magnetic field is polarized along the $x$ direction with an amplitude of $H_0$. Figure 2(a) shows the simulated normalized magnetic-field $H_x/H_0$ distributions under the illumination of TM-polarized Gaussian beams with $\theta_i = \theta_B = 55.6°$ (the left panel) and $\theta_i = -\theta_B = -55.6°$ (the right panel) at $\lambda_0 = 800$nm. Clearly, reflection is absent in both cases. Meanwhile, perfect transmission is observed under $\theta_i = \theta_B = 55.6°$, in which case, the light entering the metasurface propagates along the Cr films. While under $\theta_i = -\theta_B = -55.6°$, all incident light is absorbed. Moreover, we calculate the reflectance $R$, transmittance $T$ and absorptance $A$ as functions of the incident angle $\theta_i$ and working wavelength, as plotted in Figs. 2(b)-2(d), respectively. The dispersive parameter of Cr is taken from [41]. We see that the reflection is symmetric with respect to $\theta_i$, and is quite low ($R < 0.1$) for all angles $|\theta_i| \leq 75°$ In particular, under $\theta_i = \pm\theta_B = \pm55.6°$, we have $R < 0.01$ in the spectrum from 400-1400nm. On the other hand, the transmission and absorption show a distinct asymmetric behavior. Near-perfect transmission under $\theta_i = \theta_B = 55.6°$, and near-perfect absorption under $\theta_i = -\theta_B = -55.6°$ are observed in a wide spectrum.

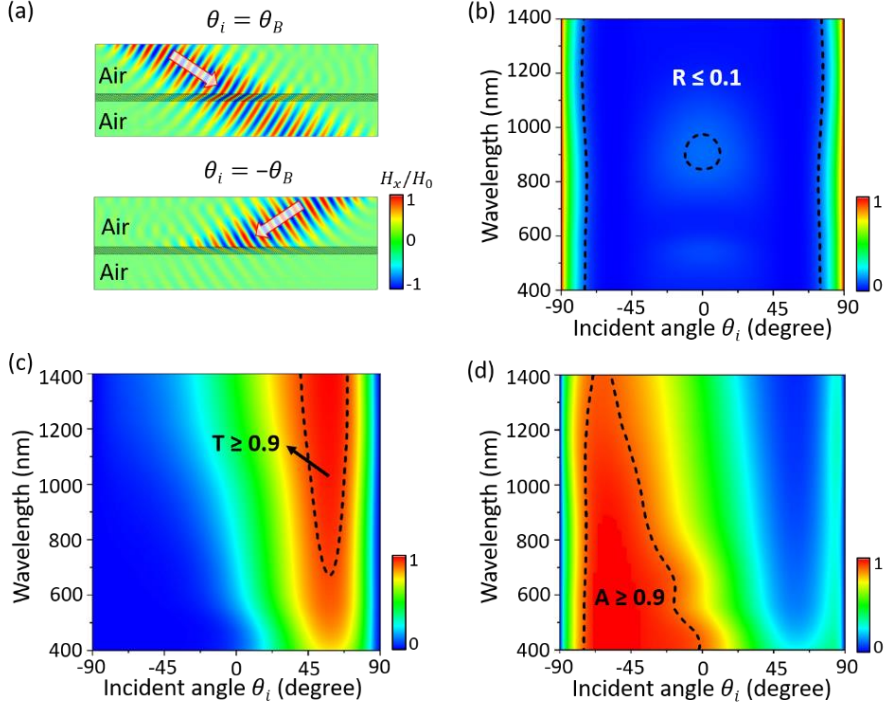

FIG. 2. (a) Simulated normalized magnetic-field $H_x/H_0$ distributions under the illumination of TM-polarized Gaussian beams with $\theta_i = \theta_B = 55.6°$ (left) and $\theta_i = -\theta_B = -55.6°$ (right) at $\lambda_0 = 800$nm. The Brewster metasurface consists of tilted Cr films in a SiO$_2$ host. The relevant parameters are $a = 100$nm, $t = 10$nm, $d = 400$nm and $\alpha = 34.4°$. (b) Reflectance, (c) transmittance, and (d) absorptance as functions of the incident angle and working wavelength.

It is noteworthy that the extreme angular-asymmetry of the optical Brewster metasurface is robust against the variations of geometrical parameters. In Figs. 3(a) and 3(b), we successively change the separation distance $a$ and thickness $t$ of Cr films. The green, blue and red lines denote, respectively, absorptance under $\theta_i = -\theta_B = -55.6°$, transmittance under $\theta_i = \theta_B = 55.6°$, and reflectance under $\theta_i = \pm\theta_B = \pm55.6°$ at $\lambda_0 = 800$nm. We can see that the near-zero reflection remains almost unchanged. High transmission due to the BE, and near-perfect absorption due to the ABE are still seen. From Fig. 3(b), one may notice that the transmission decreases as the increase of $t$. This is because the refracted light can "see" the Cr films if they are not ultra-thin. Moreover, we demonstrate an optical Brewster metasurface with random $a$ and $t$ in different transversal positions, as shown schematically in Fig. 3(c). Figure 3(d) presents the $H_x/H_0$- distributions when a TM-polarized Gaussian beam is incident onto the random metasurface under $\theta_i = \theta_B = 55.6°$ (the upper panel) and $\theta_i = -\theta_B = -55.6°$ (the lower lower), showing the occurrence of near-perfect transmission due to BE and near-perfect absorption due to the ABE. These results manifest the robustness of the extreme angular-asymmetry in the presence of imperfections of the Brewster metasurface, facilitating practical fabrication in experiments.

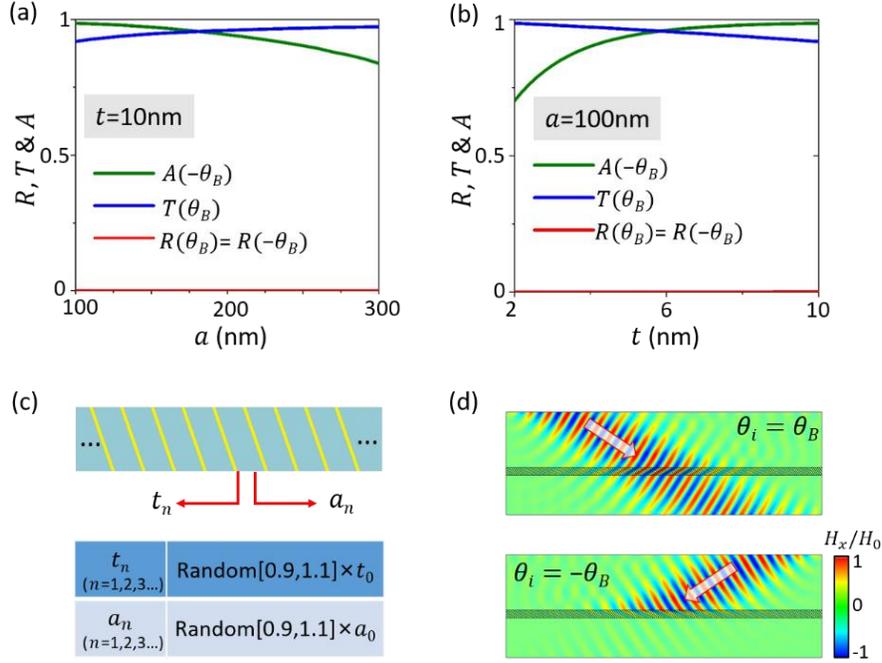

FIG. 3. [(a) and (b)] Reflectance under $\theta_i = \pm\theta_B = \pm 55.6°$, transmittance under $\theta_i = \theta_B = 55.6°$ and absorptance under $\theta_i = -\theta_B = -55.6°$ when (a) the separation distance $a$ is changed with fixed $t = 10$nm, (b) the thickness of Cr films $t$ is changed with fixed $a = 100$nm. (c) Schematic graph of an optical Brewster metasurface with random $a$ and $t$ in different transversal positions. The $a$ varies randomly in the range from $0.9a_0$ to $1.1a_0$ with $a_0 = 100$nm, and the $t$ varies randomly in the range from $0.9t_0$ to $1.1t_0$ with $t_0 = 10$nm. (d) Simulated $H_x/H_0$-distributions when a TM-polarized Gaussian beam is incident from air onto the random metasurface under $\theta_i = \theta_B = 55.6°$ (upper) and $\theta_i = -\theta_B = -55.6°$ (lower). The working wavelength in (a)-(d) is 800nm.

## III. FFECTIVE-MEDIUM DESCRIPTION AND EVALUATION OF MATERIAL SELECTION

The above results can also be understood from the perspective of effective medium model. For wavelengths much larger than the separation distance $a$, the metasurface can be approximately homogenized as an anisotropic effective medium with $\varepsilon_\perp = \frac{\varepsilon_m \varepsilon_d a \cos\alpha}{\varepsilon_m(a\cos\alpha - t) + \varepsilon_d t}$ and $\varepsilon_\parallel = \varepsilon_d + \frac{\varepsilon_m - \varepsilon_d}{a\cos\alpha} t$, which are, respectively, the effective permittivities normal and parallel to the metal films[44, 45]. Considering the limit of $t \ll a$, the $\varepsilon_\perp$ can be simplified to $\varepsilon_\perp \approx \varepsilon_d$. When light is incident under $\theta_i = \theta_B$, the electric field of the refracted light is polarized along the direction of $\varepsilon_\perp$, which is irrelevant to metal films. Therefore, for the incident light, the metal films do not seem to exist. Intriguingly, when flipping the incident angle to the anomalous Brewster's angle $-\theta_B$, the zero reflection preserves according to the reciprocity principle, but the refracted light can "see" both $\varepsilon_\perp$ and $\varepsilon_\parallel$.

Since the $\varepsilon_\parallel$ is $\varepsilon_m$-dependent, the incident light under $\theta_i = -\theta_B$ would be absorbed by metal films. Here we express the $\varepsilon_m$ in a complex form as $\varepsilon_m = \varepsilon_m' + i\varepsilon_m''$ with $\varepsilon_m'$ and $\varepsilon_m''$ being the real and imaginary parts, respectively. Generally, the real part $\varepsilon_m'$ is a negative value for Drude metals like silver and gold in the optical regime, and its absolute value becomes larger for longer wavelengths. However, a large $\varepsilon_m'$ would lead to short skin depth for light in metals, hindering the absorption of light. On the other hand, the imaginary part $\varepsilon_m''$ should be large, otherwise the imaginary part of $\varepsilon_\parallel$ would be negligibly small due to $t \ll a$. Therefore, to gain high-efficiency light absorption, the metal films should satisfy the condition $|\varepsilon_m''| \gg |\varepsilon_m'| \sim \varepsilon_d$ in a wide spectrum. This is the reason the metal Cr instead of silver and gold is chosen here. The permittivity of Cr is found to be $\varepsilon_m = -2.0 + 21.9i$ at 800nm, and approximately fulfill the condition $|\varepsilon_m''| \gg |\varepsilon_m'| \sim \varepsilon_d$ over the spectrum from 400-1400nm [41]. Besides the Cr, we find that V, W, Ti and Mo are also promising candidate metals for the optical Brewster metasurfaces.

For a demonstration, we recalculate the absorptance of the metasurface in Fig. 2 under $\theta_i = -\theta_B = -55.6°$ when different kinds of metal films are exploited instead, as plotted in Fig. 4(a). Evidently, high absorption is observed in the spectrum from 400-1400nm. Since the reflection is absent due to the ABE, the absorptance can be further increased to be near-unity through simply increasing the thickness $d$ of the metasurface. Figure 4(b) presents the absorptance as a function of $d$ at $\lambda_0 = 800$nm, showing that near-perfect absorption of light can be obtained within a thickness of one wavelength.

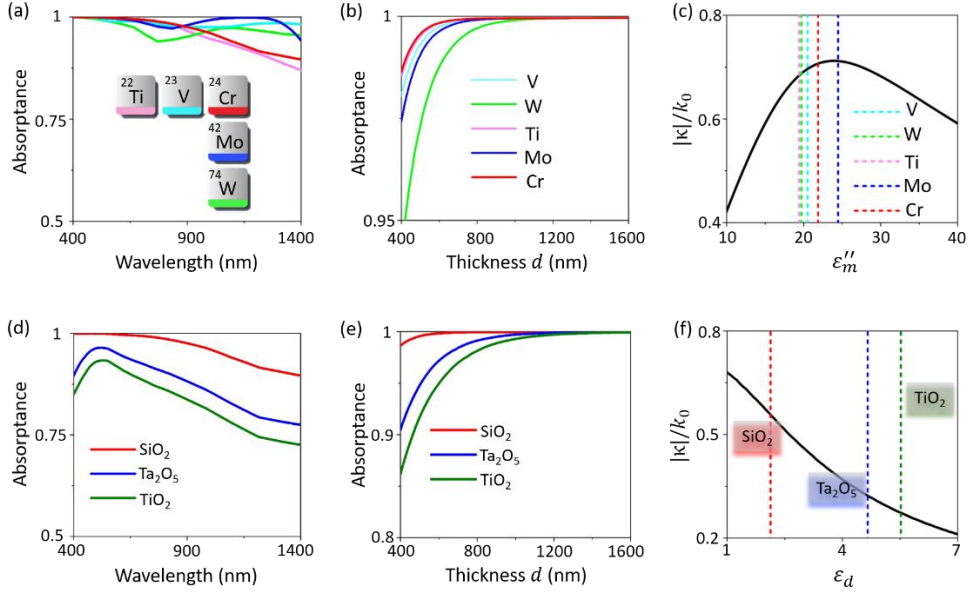

FIG. 4. (a) Absorption spectra under $\theta_i = -\theta_B = -55.6°$. The metasurface is the same as that in Fig. 2 except for the materials of metal films. (b) Absorptance under $\theta_i = -\theta_B = -55.6°$ at $\lambda_0 = 800$nm when the thickness $d$ is increased. (c) Normalized decay rate of light in the metasurface as the function of $\varepsilon_m''$ (black solid lines). The vertical dashed lines denote the values of $\varepsilon_m''$ of V, W, Ti, Mo and Cr. (d) Absorption spectra under $\theta_i = -\theta_B$. The metasurface is the same as that in Fig. 2 except for the materials of dielectric hosts. The $\theta_B$ changes accordingly for different dielectric hosts. (e) Absorptance under $\theta_i = -\theta_B$ at $\lambda_0 = 800$nm with increasing the thickness $d$. (f) Normalized decay rate as the function of $\varepsilon_d$ (black solid lines). The vertical dashed lines denote the values of $\varepsilon_d$ of $SiO_2$, $Ta_2O_5$ and $TiO_2$.

In order to quantitatively evaluate the absorption efficiency, we express the magnetic field of light inside the metasurface as $H_0 e^{i(k_y y + k_z z) - i\omega t}$, where $k_x$ ($= k_0 \sin\theta_i$) and $k_z$ are, respectively, the $x$- and $z$-components of the wave vector inside the metasurface. $k_0$ is the wave number in free space, and $\omega$ is the angular frequency. Due to the material loss, $k_z$ is a complex value, and thus can be expressed as $k_z = k'_z + i\kappa$ with $k'_z$ and $\kappa$ being the real and imaginary parts, respectively. This indicates that the refracted light decays exponentially at a rate $\kappa$. Since there is no reflection, a larger decay rate indicates the higher absorption efficiency. The $k_x$ and $k_z$ are connected by the dispersion of the effective medium as [46],

$$k_y^2\varepsilon_{yy} + k_z^2\varepsilon_{zz} + k_y k_z(\varepsilon_{yz} + \varepsilon_{zy}) = (\varepsilon_{yy}\varepsilon_{zz} - \varepsilon_{yz}\varepsilon_{zy})k_0^2 \tag{1}$$

where $\varepsilon_{yy}$, $\varepsilon_{yz}$, $\varepsilon_{zy}$ and $\varepsilon_{zz}$ are the terms of the permittivity tensor $\bar{\varepsilon} = \begin{pmatrix} \varepsilon_{yy} & \varepsilon_{yz} \\ \varepsilon_{zy} & \varepsilon_{zz} \end{pmatrix} = \begin{pmatrix} \varepsilon_\perp \cos^2\alpha + \varepsilon_\parallel \sin^2\alpha & (\varepsilon_\perp - \varepsilon_\parallel)\sin\alpha\cos\alpha \\ (\varepsilon_\perp - \varepsilon_\parallel)\sin\alpha\cos\alpha & \varepsilon_\perp \sin^2\alpha + \varepsilon_\parallel \cos^2\alpha \end{pmatrix}$. Considering the condition $|\varepsilon''_m| \gg |\varepsilon'_m| \sim \varepsilon_d$, the $\varepsilon_\parallel$ can be simplified to $\varepsilon_\parallel \approx \varepsilon_d + i\gamma$ with $\gamma = \frac{\varepsilon''_m}{a\cos\alpha}t$. Separating the real and imaginary parts of Eq. (1), and eliminating the variable $k'_z$ yields

$$\left(\frac{\kappa}{k_0}\right)^3 [\gamma^2 + (1+\varepsilon_d)^2]^2 + 4\left(\frac{\kappa}{k_0}\right)^2 \gamma\sqrt{1+\varepsilon_d}[\gamma^2 + (1+\varepsilon_d)^2] + \left(\frac{\kappa}{k_0}\right)(1+\varepsilon_d)[\gamma^4 + \varepsilon_d^2(1+\varepsilon_d)^2 + \gamma^2(5 + 2\varepsilon_d + 2\varepsilon_d^2)] + 2\gamma(1+\varepsilon_d)^{3/2}(\gamma^2 + \varepsilon_d^2) = 0 \tag{2}$$

Based on Eq. (2), we plot the normalized decay rate $|\kappa|/k_0$ with respect to $\varepsilon''_m$ (the black solid lines in Fig. 4(c)), showing a maximal decay rate occurring at $\varepsilon''_m = 23.92$. The vertical dashed lines denote the values of $\varepsilon''_m$ of V, W, Ti, Mo and Cr. We find that all these values are close to the optimal value, i.e. $\varepsilon''_m = 23.92$. This further confirms that V, W, Ti, Mo and Cr are excellent candidate metals for high-efficiency optical Brewster metasurfaces.

Besides the metals, the dielectric hosts also affect the absorption efficiency as both $\varepsilon_\perp$ and $\varepsilon_\parallel$ are $\varepsilon_d$-dependent. To explore the influences, we take high-index tantalum pentoxide (Ta$_2$O$_5$) and titanium dioxide (TiO$_2$) for comparison, whose refractive indices are 2.16 and 2.35, respectively. Figure 4(d) presents the absorption spectra of the metasurface in Fig. 2 under the anomalous Brewster's angle $\theta_i = -\theta_B = -\arctan\sqrt{\varepsilon_d}$ when different kinds of dielectric hosts are considered. Since the $\theta_B$ relies on $\varepsilon_d$, the incident angle is changed accordingly for different dielectric hosts. It is seen that the absorption decreases when the high-index host is utilized. Through increasing the thickness $d$, near-perfect light absorption can still be obtained, as shown in Fig. 4(e). This indicates that the imperfection of absorption in Fig. 4(d) attributes to the small decay rate, rather than the reflection. For further verification, we plot the normalized decay rate $|\kappa|/k_0$ as a function of $\varepsilon_d$, as presented by the black solid lines in Fig. 4(f). It is clearly seen that the decay rate decreases as the increase of $\varepsilon_d$. Since the SiO$_2$ possesses the smallest refractive index among the three dielectrics, the absorption efficiency is the highest, as observed in Fig. 4(d).

The material selection of dielectric hosts affects not only the absorption efficiency, but also the absorption bandwidth. To gain ultra-broadband perfect absorption of light, the anomalous Brewster's angle should be a constant value over an ultra-wide spectrum, which requires a dispersion-less $\varepsilon_d$. Within the spectrum of dispersion-less $\varepsilon_d$, there is no extra restriction on the lower limit of working frequency, while the upper limit is determined by the validity of the effective medium approximation. Therefore, we can conclude that the low-index dielectrics possessing low chromatic dispersions like SiO$_2$ and MgF$_2$ are promising candidate materials for the ultra-broadband and high-efficiency optical Brewster metasurfaces.

## IV. STRATEGIES TO IMPROVE ABSORPTION PERFORMANCE

In the following, we show that the perfect absorption under the anomalous Brewster's angle $-\theta_B$ can be extended to the traditional Brewster's angle $\theta_B$, so as to greatly improve the absorption performance. The first approach is to integrate a metal back-reflector. As illustrated in Fig. 5(a), when light is incident under $\theta_i = \theta_B$, all light can transmit into the metasurface because of the BE. Due to the back-reflector, the transmitted light will be reflected and then completely absorbed by the metal films. We note that there is no restriction on the choice of metals for the back-reflector. For simplicity, we utilize a bottom Cr layer to serve as the back-reflector. Figure 5(b) presents the simulated results for TM-polarized light under $\theta_i = \theta_B = 55.6°$ (the left panel) and $\theta_i = -\theta_B = -55.6°$ (the right panel) at $\lambda_0 = 800$nm, showing near-perfect light absorption in both cases. The relevant parameters are $a = 100$nm, $t = 10$nm, $d = 800$nm and $\alpha = 34.4°$. Moreover, the absorptance as functions of the incident angle $\theta_i$ and working wavelength is calculated and plotted in Fig. 5(c). We see that the absorption becomes symmetric with respect to $\theta_i$, and is significantly improved. High absorption ($A > 0.9$) is obtained for all angles $|\theta_i| \leq 75°$ in the spectrum from 400-1400nm.

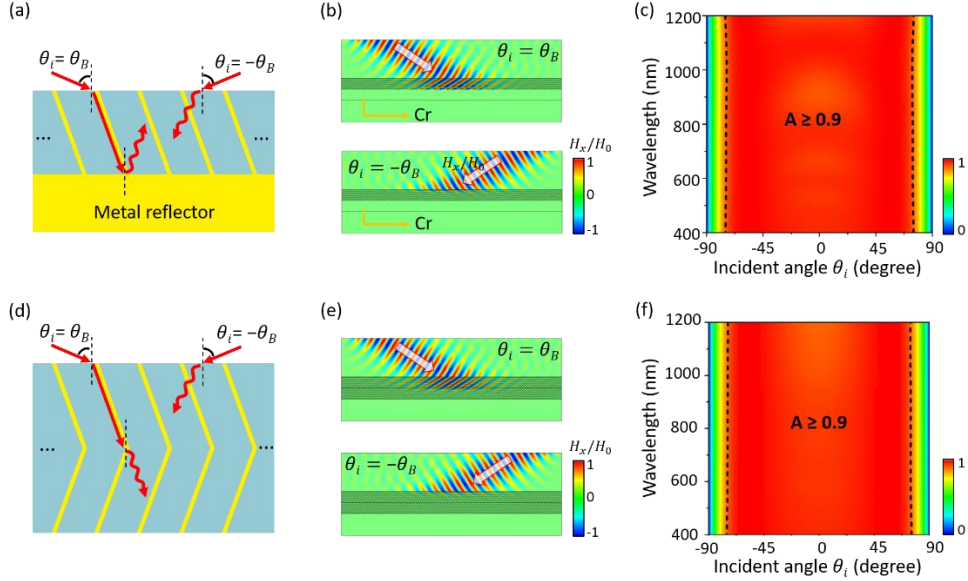

FIG. 5. [(a) and (d)] Illustration of an optical Brewster metasurface with (a) a metal back-reflector, (d) a folded metasurface. [(b) and (e)] Simulated $H_x/H_0$-distributions a TM-polarized Gaussian beam is incident under $\theta_i = \theta_B = 55.6°$ (left) and $\theta_i = -\theta_B = -55.6°$ (right) at $\lambda_0 = 800$nm. The relevant parameters are $a = 100$nm, $t = 10$nm, $d = 800$nm and $\alpha = 34.4°$. In (b), the back-reflector is made of Cr. In (e), the bottom half is a folded metasurface. [(c) and (f)] Absorptance as functions of the incident angle and working wavelength regarding to the metasurfaces in (b) and (e), respectively.

The second approach is to integrate a folded Brewster metasurface. As illustrated in Fig. 5(d), the top half is the same as that in Fig. 5(a), while the bottom half is a folded metasurface in which the tilt angle of metal films is flipped to $-\alpha$. In such a configuration, the incident light under $\theta_i = \theta_B$ can totally transmit through the top half and then be dissipated in the bottom half. For verification, we simulate the $H_x/H_0$-distributions for TM-polarized Gaussian beams with $\theta_i = \pm\theta_B = \pm55.6°$ at $\lambda_0 = 800$nm (Fig. 5(e)). As expected, all incident light is absorbed in both cases. In addition, the absorptance with respect to $\theta_i$ and working wavelength (Fig. 5(f)) indicates that the absorption performance of the metasurface is indeed greatly improved.

We note that the angle range of high absorption can be further enlarged by exploiting high-index dielectric hosts, because both the traditional and anomalous Brewster's angles become larger. However, this would lead to the decrease of absorption efficiency, as discussed in Figs. 4(d)-4(f), indicating that a larger thickness of the metasurface is required to absorb the same amount of incident energy. Therefore, there is a trade-off between the angle range and absorption efficiency (or thickness) in the improvement of absorption performance.

## V. GRADIENT OPTICAL BREWSTER METASURFACES FOR ULTRA-BROADBAND AND NEAR-OMNIDIRECTIONAL ABSORPTION

The above discussions suggest that through harnessing the tilt angle of metal films inside the Brewster metasurface, reflectionless absorption over an ultra-wide spectrum can be realized under a particular incident angle, i.e. the anomalous Brewster's angle, for light incident from free space. In the following, we'd like to show that this extraordinary absorption can be extended to all incident angles when the light source and metasurface are in the same dielectric host.

As shown schematically in Fig. 6(a), we consider a point source and a gradient metasurface in the same dielectric host. First, we assume that each metal film is aligned along the direction of emitted light from the point source, satisfying the condition $\alpha = \theta_i$ everywhere (the upper panel). In this case, all emitted light can transmit through the gradient metasurface without any reflection, similar to the traditional BE. Then, we flip the rotation angle from $\alpha$ to $-\alpha$ of each metal film, thus a gradient metasurface satisfying the condition $\alpha = -\theta_i$ everywhere is constructed (the lower panel). Considering the principle of reciprocity, there will be no reflection, like the ABE. While the light entering the metasurface would be dissipated on the metal films, bestowing omnidirectional reflectionless absorption.

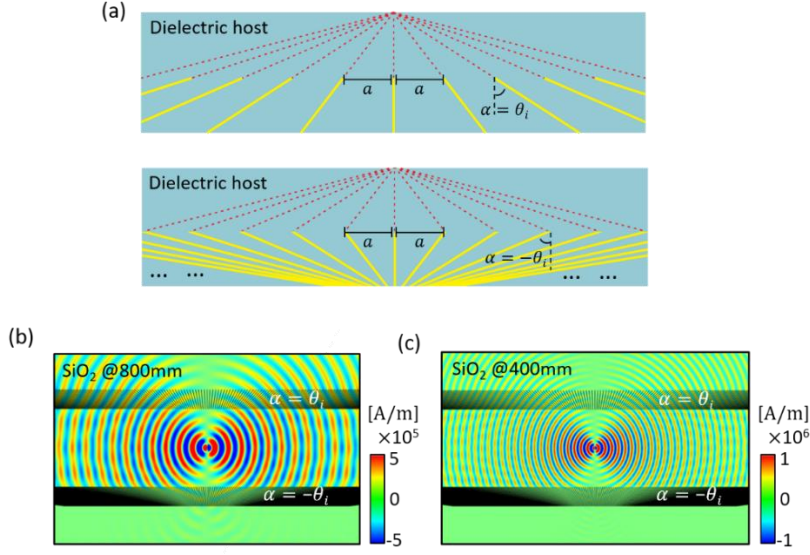

FIG. 6. (a) Schematic graphs of gradient metasurfaces with varied tilt angles of metal films satisfying $\alpha = \theta_i$ (upper) or $\alpha = -\theta_i$ (lower) everywhere in a dielectric host. [(b) and (c)] Simulated magnetic-field distributions when an electric dipole source is placed between two metasurfaces in the SiO$_2$ host at (b) $\lambda_0 = 800$nm, (c) $\lambda_0 = 400$nm. The upper (lower) metasurface satisfies the condition $\alpha = \theta_i$ ($\alpha = -\theta_i$) everywhere.

The numerical verification is performed through examining the radiation of a vertical electric dipole source (dipole moment 1A·m) between two Brewster metasurfaces in the SiO$_2$ host (Fig. 6(b) and 6(c)). Both metasurfaces consist of ultra-thin Cr films with $a = 100$nm, $t = 10$nm and $d = 800$nm. The rotation angle satisfies $\alpha = \theta_i$ ($\alpha = -\theta_i$) everywhere in the upper (lower) metasurface. Figures 6(b) and 6(c) show the simulated magnetic-field distributions at $\lambda_0 = 800$nm and $\lambda_0 = 400$nm, respectively. The well-defined dipole radiation patterns without clear interference patterns induced by reflection indicate the near-omnidirectional zero reflection on both metasurfaces. Meanwhile, the magnetic field is almost unaltered above the upper metasurface, but quite weak below the lower metasurface. These results clearly demonstrate the extraordinary phenomenon of ultra-broadband and near-omnidirectional reflectionless transmission (or absorption) of light in gradient Brewster metasurfaces through simply engineering the tilt angle of metal films. In principle, this approach is applicable for the reflectionless manipulation of light of arbitrary wavefront.

## VI. DISCUSSION AND CONCLUSION

The hallmark advantage of the proposed optical Brewster metasurfaces is that the reflectionless characteristic is mostly determined by the dielectric host and independent of the metal films. Due to the low chromatic dispersion of dielectric permittivity, the bandwidth of reflectionless absorption in principle can cover an ultra-broad spectrum, far beyond those of other absorber techniques. For example, tilted hyperbolic metamaterials using metal-dielectric multilayer composites have been proposed to realize perfect light absorption [34, 47-49]. Nevertheless, the absorption bandwidth is limited by the strong dispersion of their metal components in the visible spectral range, and thus is inherently narrow.

The extreme angular-asymmetry is another intriguing property of the optical Brewster metasurfaces. This extraordinary phenomenon is originated in the coexistence of the traditional BE and ABE occurring at the corresponding incident angles with opposite signs, and this is strictly protected by the reciprocity principle. Compared with the newly emerging schemes based on resonant metasurfaces and meta-gratings [35-37], our scheme has advantages in simplicity, high efficiency, robustness and ultra-broad bandwidth.

The principle of optical Brewster metasurfaces is universal. Besides microwaves, visible and near-infrared frequencies, it can be easily extended to far-infrared and terahertz regimes, where ultra-thin graphene might be excellent candidate materials. Due to the tunability of graphene [50, 51], dynamically controllable ABE and absorption could become possible.

In summary, we have demonstrated a simple but highly efficient approach to realize ultra-broadband reflectionless optical Brewster metasurfaces. Near-perfect absorption of light can be obtained over the entire visible spectral range and the near-infrared regime. The optical Brewster metasurfaces exhibit extreme angular-asymmetry, based on which a transition from perfect transparency to perfect absorption is achieved due to the coexistence of the traditional BE and ABE occurring at the incident angles with opposite signs. We have systematically and quantitatively evaluated the material selection guidelines for high-efficiency absorption. Strategies like the utilization of a metal back-reflector or a folded metasurface have been proposed to remove the angular asymmetry in absorption and significantly improve the performance. A gradient optical Brewster metasurface exhibiting ultra-broadband and near-omnidirectional reflectionless absorption is also shown. Our findings offer a path towards high-efficiency light absorption with ultra-broad bandwidth and extreme angular-asymmetry.

**ACKNOWLEDGEMENTS**

National Key R&D Program of China (Grant Nos. 2020YFA0211300, 2017YFA0303702) National Natural Science Foundation of China (Grant No. 11974176), the Research Grants Council of Hong Kong (Grant No. R6015-18), the Priority Academic Program Development of Jiangsu Higher Education Institutions (PAPD).